\documentclass[12pt]{article}
\usepackage{graphicx,amssymb,amsfonts}
\newlength{\extraspace}
\newlength{\extraspaces}
\newcounter{savefootnote}
\begingroup
\endgroup

\newcommand{\be}{\begin{equation}
\addtolength{\abovedisplayskip}{\extraspaces}
\addtolength{\belowdisplayskip}{\extraspaces}
\addtolength{\abovedisplayshortskip}{\extraspace}
\addtolength{\belowdisplayshortskip}{\extraspace}}
\newcommand{\ee}{\end{equation}}

\newcommand{\ba}{\begin{eqnarray}
\addtolength{\abovedisplayskip}{\extraspaces}
\addtolength{\belowdisplayskip}{\extraspaces}
\addtolength{\abovedisplayshortskip}{\extraspace}
\addtolength{\belowdisplayshortskip}{\extraspace}}
\newcommand{\ea}{\end{eqnarray}}

\newcommand{\nonu}{\nonumber \\[.5mm]}
\newcommand{\A}{&\!\!\!}

\newcommand{\newsection}[1]{
\vspace{7mm} \pagebreak[3] \addtocounter{section}{1}
\setcounter{subsection}{0} 
\begin{center}
{\large {\bf \thesection. #1}}
\end{center}
\nopagebreak
\medskip
\nopagebreak \hspace{3mm}}

\begin{document}

\begin{center}
{\bf  A special exact spherically symmetric solution in f(T) gravity theories}\footnote{ 
\hspace*{.5cm}
 }
\end{center}
\begin{center}
{\bf Gamal G.L. Nashed}
\end{center}

\bigskip

\centerline{\it Centre  for Theoretical Physics, The British
University in  Egypt} \centerline{\it Sherouk City 11837, P.O. Box
43, Egypt \footnote{ Mathematics Department, Faculty of Science, Ain
Shams University, Cairo, 11566, Egypt \\
\hspace*{.2cm} Egyptian Relativity Group (ERG) URL:
http://www.erg.eg.net}}

\bigskip
\centerline{ e-mail:nashed@bue.edu.eg}

\hspace{2cm} \hspace{2cm}
\\
\\
\\
\\
\\

A non-diagonal spherically symmetric tetrad field, involving
 four unknown functions of radial coordinate $r$, is applied to the equations
 of motion  of f(T) gravity theory. A special exact   vacuum solution
  with one constant of integration
is obtained. The scalar torsion related to this special solution
vanishes. To understand the physical meaning of the constant of
integration we calculate the energy associated with this solution
and show how it is related to the gravitational mass of the system.

\begin{center}
\newsection{\bf Introduction}
\end{center}

Common consensus in the scientific circle is that the
characterization of the gravitational field, powered by Einstein
General Relativity  (GR), is  inaccurate at scales of  magnitude
of the Planck$'s$ length.  The spacetime frame of such
characterization should be clarified by a quantum regime. In the
opposite extreme  of the physical phenomena, GR also faces a
fascinating problem linked to the late cosmic speed-up stage of the
universe. According to the previous reasons, and for other defects,
GR has been the topic of many modifications which have attempted to
supply a more satisfactory qualification  of the gravitational field
in the above aforementioned extreme regimes. Among the most
important modified gravitational theories  is the one called
``$f(T)$ gravity", which is a theory constructed in a spacetime
having absolute parallelism (cf., \cite{HHK}-\cite{HS9}).

 Many of $f(T)$ gravity theories have been analyzed in \cite{BF9}-\cite{WY1}. It
 has been suggested that $f(T)$ gravity theory is not dynamically synonymous
with the teleparallel equivalent of GR  Lagrangian through conformal
transformation \cite{Yr1}. Many observational restrictions have been
studied  \cite{Bg1}-\cite{WMQ}. Large-scale structure in $f(T)$
gravity theory has been analyzed \cite{LSB}. Perturbations in the
area of cosmology in $f(T)$ gravity have been demonstrated
\cite{DDS}-\cite{CCDDS}. Birkhoff$'s$ theorem,
 in $f(T)$ gravity has  been studied \cite{MW1}. Stationary, spherical symmetry solutions  have
been derived for  $f(T )$ theories \cite{Wt1}. Relativistic stars
and the cosmic expansion have been investigated \cite{Dy,SSFR}.

$f(T)$ gravity theories have engaged many concerns: It has been
indicated that the Lagrangian  and the  equations of motion  are not
 invariant under local Lorentz transformations \cite{LSB1}. The reasons
  for  such phenomena  has been explained in \cite{SLB1}.

The equations of motion of $f(T)$  theories
differ from those of $f(R)$ theories \cite{NO}$-$\cite{BMT}, because
they are of the second order instead of  the fourth order as in
$f(R)$ theories. Such property has been believed as  an indicator
which shows that the theory might be of much interest. The non locality
of such theories indicate that $f(T)$  seems  to comprise more
degrees of freedom.

The target of this study is to find an analytic vacuum spherically
symmetric solution, within the framework of higher-torsion theories,
i.e., for $f(T)$ gravitational theory. In \S 2, a brief review is
presented of the covariant formalism for the gravitational
energy-momentum, described by the pair $( \vartheta^{\alpha},
{\Gamma_\alpha}^\beta)$.
 In \S 3, a brief survey of the $f(T)$ gravitational theory is provided.
 A non-diagonal spherically symmetric tetrad field, with four
 unknown functions of the radial coordinate $r$, is
presented. The application  of such tetrad field to the  equations
of motion of  $f(T)$ is demonstrated in \S 4. Also, in \S 4, an
analytic vacuum spherically symmetric solution with one constant of
integration is obtained.  In \S 5,  we calculate the energy
associated with this solution in order to understand the physical
meaning of the constant. This calculation uses traditional
computation employing the  Riemannian connection:
${\Gamma_\alpha}^\beta$. The
 final section is devoted to discussions of these results.
\\
\\
\\
\centerline{\bf Notation}

By convention, we denote the exterior vector product  by $\wedge$,
 the interior  of a vector product $\xi$ and a p-form $\Psi$  by
$\xi \rfloor \Psi$. The vector basis dual to the frame 1-forms
$\vartheta^{\alpha}$ is denoted by $e_\alpha$ and they satisfy
$e_\alpha \rfloor \vartheta^{\beta}={\delta_\alpha}^\beta$. Using
local coordinates $x^i$, we have $\vartheta^{\alpha}=h^\alpha_i
dx^i$ and $e_\alpha=h^i_\alpha
\partial_i$ where $h^\alpha_i$ and $h^i_\alpha $ are the covariant
and contravariant components of the tetrad field. We define the
volume 4-form by \be \eta \stackrel {\rm def.}{=}
\vartheta^{\hat{0}}\wedge \vartheta^{\hat{1}}\wedge
\vartheta^{\hat{2}}\wedge\vartheta^{\hat{3}}.\ee  Furthermore, with
the help of the interior product we define \[\eta_\alpha \stackrel
{\rm def.}{=} e_\alpha \rfloor \eta = \ \frac{1}{3!} \
\epsilon_{\alpha \beta \gamma \delta} \ \vartheta^\beta \wedge
\vartheta^\gamma \wedge
\vartheta^\delta={^*\vartheta}_{\alpha}\footnote{\textrm \ \
* \ \ refers \ \ to \ \ Hodge \ \ dual \ \ operator,},
\]  where
$\epsilon_{\alpha \beta \gamma \delta}$ is completely antisymmetric
with $\epsilon_{0123}=1$. \[\eta_{\alpha \beta} \stackrel {\rm
def.}{=} e_\beta \rfloor \eta_\alpha = \frac{1}{2!}\epsilon_{\alpha
\beta \gamma \delta} \ \vartheta^\gamma \wedge
\vartheta^\delta={^*\left(\vartheta_\alpha \wedge
\vartheta_\beta\right)},\]
\[\eta_{\alpha \beta \gamma} \stackrel {\rm def.}{=} e_\gamma
\rfloor \eta_{\alpha \beta}= \frac{1}{1!} \epsilon_{\alpha \beta
\gamma \delta} \ \vartheta^\delta={^*\left(\vartheta_\alpha \wedge
\vartheta_\beta\wedge \vartheta_\gamma\right)},\]  which are bases
for 3-, 2- and 1-forms respectively.

Finally, \[\eta_{\alpha \beta \mu \nu} \stackrel {\rm def.}{=} e_\nu
\rfloor \eta_{\alpha \beta \mu}= e_\nu \rfloor e_\mu \rfloor e_\beta
\rfloor e_\alpha \rfloor \eta={^*\left(\vartheta_\alpha \wedge
\vartheta_\beta \wedge \vartheta_\mu \wedge \vartheta_\nu\right)},\]
is the Levi-Civita tensor density. The $\eta$-forms satisfy the
useful identities: \ba \vartheta^\beta \wedge \eta_\alpha \A=
 \A \delta^\beta_\alpha \eta, \qquad
\vartheta^\beta \wedge \eta_{\mu \nu}  = \delta^\beta_\nu
\eta_\mu-\delta^\beta_\mu \eta_\nu, \qquad \vartheta^\beta \wedge
\eta_{\alpha \mu \nu}  =\delta^\beta_\alpha \eta_{\mu
\nu}+\delta^\beta_\mu \eta_{\nu \alpha}+\delta^\beta_\nu \eta_{
\alpha \mu}, \nonu
\vartheta^\beta \wedge \eta_{\alpha \gamma \mu \nu}  \A =  \A
\delta^\beta_\nu \eta_{\alpha \gamma \mu}-\delta^\beta_\mu
\eta_{\alpha \gamma \nu }+\delta^\beta_\gamma \eta_{ \alpha \mu
\nu}-\delta^\beta_\alpha \eta_{ \gamma \mu \nu}. \ea
\newsection{Brief review of teleparallel gravity}
Teleparallel geometry can be viewed as a gauge theory of translation
\cite{Hw}-\cite{Tr}. The coframe $\vartheta^\alpha$  plays the role
of the gauge translational potential of the gravitational field.  GR
can be reformulated as a teleparallel theory. Geometrically,
teleparallel gravity can be considered as a special case
 of the Metric-Affine Gravity (MAG) in which
  $\vartheta^\alpha$ and the local Lorentz connection
 are subject to the distant parallelism
constraint ${R_\alpha}^\beta=0$ (cf., \cite{OP1}-\cite{Oy}). In this
geometry the torsion 2-form  \be
T^\alpha=D\vartheta^\alpha=d\vartheta^\alpha+{\Gamma_\beta}^\alpha\wedge
\vartheta^\beta=\frac{1}{2}{T_{\mu \nu}}^\alpha \vartheta^\mu \wedge
\vartheta^\nu=\frac{1}{2}{T_{i j}}^\alpha dx^i \wedge dx^j,\ee
arises as the gravitational gauge field strength,
${\Gamma_\alpha}^\beta$ being the Weitzenb\"ock 1-form connection,
$d$ in front of the coframe $\vartheta^\alpha$  the exterior
derivative and $D$ the covariant derivative associated with
${\Gamma_\alpha}^\beta$.

The torsion form $T^\alpha$ can be decomposed into three irreducible
pieces \cite{HS, LOP}: the tensor, the trace and the axial trace
given respectively by  \ba {^ {\tiny{( 1)}}T^\alpha} \A \stackrel
{\rm def.}{=} \A T^\alpha-{^ {\tiny{( 2)}}T^\alpha}-{^ {\tiny{(
3)}}T^\alpha}, \qquad {\textrm with} \nonu
{^  {\tiny{( 2)}}T^\alpha}  \A \stackrel {\rm def.}{=} \A
\frac{1}{3} \vartheta^\alpha\wedge T, \quad {\textrm where} \quad T=
\left(e_\beta \rfloor T^\beta\right), \qquad e_\alpha \rfloor
T={T_{\mu \alpha}}^\mu, \quad {\textrm vector \ trace \  of \ torsion}
\nonu
{^  {\tiny{( 3)}}T^\alpha}  \A \stackrel {\rm def.}{=} \A
\frac{1}{3} e^\alpha\rfloor P, \quad {\textrm with} \quad
P=\left(\vartheta^\beta \wedge T_\beta\right), \quad e_\alpha\rfloor
P=T^{\mu \nu \lambda}\eta_{\mu \nu \lambda \alpha}, \quad {\textrm
axial \ trace \ of \ torsion}.\nonu
\A \A \ea The Lagrangian of the teleparallel equivalent of GR has
the form \cite{LOP}\footnote{The effect of adding the non- Riemannian parity
odd pseudoscalar curvature to the Hilbert-Einstein-Cartan scalar
curvature was studied by many authors (cf., \cite{M09} and
references therein).} \be V( \vartheta^{\alpha},
{\Gamma_\alpha}^\beta)= -\frac{1}{2\kappa}T^\alpha \wedge
{^*\left({^ {\tiny{( 1)}}T_\alpha}-2{^  {\tiny{( 2)}}T_\alpha}
-\frac{1}{2}{^ {\tiny{( 3)}}T_\alpha} \right)}, \ee where
$\kappa=8\pi G/c^3$, $G$ is the Newtonian constant, $c$ is the speed
of light and the metric $g_{\alpha \beta}$
 is assumed to be flat Minkowski metric $g_{\alpha
\beta}=O_{\alpha \beta}=diag(+1,-1,-1,-1)$, that is used to raise
and lower local frame (Greek) indices. In accordance with the
general Lagrange-Noether scheme \cite{HMMN,Gf}, one can derive from
equation  (5) the translational momentum 2-form $H_{\alpha}$ and the
canonical energy-momentum 3-form $E_\alpha$ are given respectively
by: \be H_{\alpha}=-\frac{\partial V}{\partial
T^\alpha}=\frac{1}{\kappa} {^\ast\left({^ {\tiny{( 1)}}T_\alpha}-2{^
{\tiny{( 2)}}T_\alpha} -\frac{1}{2}{^ {\tiny{(
3)}}T_\alpha}\right)}, \ee \be E_\alpha \stackrel {\rm def.}{=}
\frac{\partial V}{\partial \vartheta^\alpha}=e_\alpha \rfloor
V+\left(e_\alpha \rfloor T^\beta \right) \wedge H_\beta. \ee Due to
geometric identities \cite{Oyn}, the Lagrangian of equation (5) can
be rewritten  as \be V=-\frac{1}{2}T^\alpha\wedge H_\alpha.\ee

The variation of the total action with respect to the coframe gives
the  equations of motion in the  form:  \be
DH_\alpha-E_\alpha=\Sigma_\alpha, \ \ {\textrm where} \ \
\Sigma_\alpha \stackrel {\rm def.}{=} \frac{\delta
L_{mattter}}{\delta \vartheta^\alpha},\ee  is the canonical
energy-momentum current 3-form   which is considered as the source
of matter. The presence of the connection form
$\Gamma^\alpha_\beta$ in equation (3) plays an important
regularizing
role due to the following: \vspace{.3cm}\\
\underline{First}: The theory becomes explicitly covariant under the
local Lorentz transformations of the coframe, i.e.
 the special Lagrangian given by equation  (5) is invariant under the change of
variables
 \be   \vartheta'^\alpha={L^\alpha}_\beta
\vartheta^\beta, \qquad
{\Gamma'_\alpha}^\beta={\left(L^{-1}\right)^\mu}_\alpha
{\Gamma_\mu}^\nu {L^\beta}_\nu+{L^\beta}_\gamma
d{\left(L^{-1}\right)^\gamma}_\alpha.\ee  When
${\Gamma_\alpha}^\beta=0$, which is the tetrad gravity,
  the Lagrangian (5) is only quasi-invariant, i.e., it changes by a total
divergence.\vspace{0.5cm}\\
 \underline{Second}: ${\Gamma_\alpha}^\beta$ plays an
important role in the teleparallel framework.  This role represents
the inertial effects which arise from the choice of the reference
system \cite{ALP}. The  contributions of this inertial in many cases
lead to a non-physical results for the total energy of the system.
Therefore, the role of the teleparallel connection is to separate
the inertial contribution from the truly gravitational one. Since
the teleparallel curvature is zero, the connection is a ``pure
gauge", that is \be {\Gamma_\alpha}^\beta\stackrel {\rm
def.}{=}{{\left(\Lambda^{-1}\right)}^\beta}_\gamma
d\left({\Lambda^\gamma}_\alpha\right).\ee The  Weitzenb\"ock
connection form ${\Gamma_\alpha}^\beta$ always has the form of
equation (11).

The   translational momentum of the Lagrangian (5) has the form
\cite{LOP}
 \be
\widetilde{H}_\alpha=\frac{1}{2\kappa}{\widetilde{\Gamma}}^{\beta
\gamma}\wedge  \eta_{\alpha \beta \gamma}, \qquad {\textrm where}
\qquad {\Gamma_\alpha}^\beta \stackrel {\rm def.}{=} {\tilde
{\Gamma_\alpha}}^\beta -{K_\alpha}^\beta,\ee  with ${\tilde
{\Gamma_\alpha}}^\beta $  is the purely Riemannian connection form
and $K^{\mu \nu}$ is the contortion 1-form which is related to the
torsion through the relation \be T^\alpha  \stackrel {\rm def.}{=}
{K^\alpha}_\beta \wedge \vartheta^\beta.\ee
\newsection{Brief review of f(T)}
In a spacetime having  absolute parallelism the parallel vector
field ${h_\mu}^i$ \cite{Wr} defines the nonsymmetric affine
connection: \be {\Gamma^i}_{j k} \stackrel{\rm def.}{=} {h_\mu}^i
{h^\mu}_{j, k}, \ee where $h_{\mu i, \ j}=\partial_j h_{\mu
i}$\footnote{We use the Latin indices ${\it i, j, \cdots }$ for
local holonomic spacetime coordinates and the Greek indices
$\alpha$, $\beta$, $\cdots$ for the (co)frame components}. The
curvature tensor of ${\Gamma^i}_{j  k}$ vanishes identically. The
metric tensor $g_{i j}$
 is defined by
 \be g_{i j} \stackrel{\rm def.}{=}  O_{ \mu \nu} {h^\mu}_i {h^\nu}_j. \ee
 Define the torsion components by: \ba \A \A {T^i}_{j k}   \stackrel
{\rm def.}{=}  {\Gamma^i}_{k j}-{\Gamma^i}_{j k} ={h_\mu}^i
\left(\partial_j{h^\mu}_k-\partial_k{h^\mu}_j\right),\nonu
\A \A \hspace*{-4.5cm}{\textrm and \ the \ contortion \ components \
by:} \nonu
\A \A  {K^{i j}}_k  \stackrel {\rm def.}{=}  -\frac{1}{2}\left({T^{i
j}}_k-{T^{j i}}_k-{T_k}^{i j}\right). \ea Here the contortion equals
the difference between Weitzenb\"ock and Levi-Civita connections,
i.e., ${K^{i}}_{j k}= {\Gamma^i}_{j k }-\left \{_{j k}^i\right\}$.

The tensor ${S_i}^{j k}$ is defined as \be {S_i}^{j k} \stackrel
{\rm def.}{=} \frac{1}{2}\left({K^{j k}}_i+\delta^j_i{T^{a
k}}_a-\delta^k_i{T^{a j}}_a\right). \ee The torsion scalar is
defined as \be T \stackrel {\rm def.}{=} {T^i}_{j k} {S_i}^{j k}.
\ee As was done in the development of  the $f(R)$ theory
\cite{NO}$-$\cite{BMT}, one can define the action of $f(T )$ theory
as \be {\cal L}({h^\mu}_i, \Phi_A)=\int d^4x
h\left[\frac{1}{16\pi}f(T)+{\cal L}_{Matter}(\Phi_A)\right], \quad
\textrm{where} \quad h=\sqrt{-g}=det\left({h^\mu}_i\right).\ee We
have assumed that the units such that $G = c = 1$. The variables
$\Phi_A$ are the matter fields.

 Considering the equation (19) as a function of the
field variables ${h^\mu}_i$, $\Phi_A$, and equating  the variation
of the function with respect to the tetrad field ${h^\mu}_i$ to zero, one
can obtain the following equation of motion \cite{BF}: \be {S_i}^{a
j} T_{,a} \
f(T)_{TT}+\left[h^{-1}{h^\mu}_i\partial_b\left(h{h_\mu}^a {S_a}^{b
j}\right)-{T^a}_{b i}{S_a}^{j b}\right]f(T)_T-\frac{1}{4}\delta^j_i
f(T)=4\pi {\cal T}^j_i,\ee where $T_{,a}=\frac{\partial T}{\partial
x^a}$, $f(T)_T=\frac{\partial f(T)}{\partial T}$,
$f(T)_{TT}=\frac{\partial^2 f(T)}{\partial T^2}$ and ${\cal
T}^\nu_\mu$ is the energy momentum tensor. In this work we are
interested in  the vacuum case of $f(T)$ theory, i.e., ${\cal
T}^j_i=0$.
\newsection{Spherically symmetric solution in  f(T) gravity theory}
Assume that the manifold possesses a stationary and spherical
symmetry with local Lorentz transformations the tetrad field $\left(
{h^\mu}_a \right)$ has the form: \be \left( {h^\mu}_a \right)=
\left( \matrix{A(r) & B(r)& 0 & 0\vspace{3mm} \cr C(r)\sin\theta
\cos\phi &D(r)\sin\theta \cos\phi&r\cos\theta \cos\phi &
-r\sin\theta \sin\phi \vspace{3mm} \cr C(r)\sin\theta
\sin\phi&D(r)\sin\theta \sin\phi&r\cos\theta \sin\phi & r\sin\theta
\cos\phi \vspace{3mm} \cr C(r)\cos\theta &
D(r)\cos\theta&-r\sin\theta & 0 \cr } \right),\ee
 where $A(r)$, $B(r)$, $C(r)$ and  $D(r)$ are four unknown functions of the radial coordinate, $r$.

    Using equations  (19) and
  (21), one can obtain $h = det ({h^\mu}_a)
  = r^2\sin \theta (AD-BC)$\footnote{We will denote $A$, $B$,
  $\cdots$ instead of $A(r)$, $A(r)$,  $\cdots$
  .}:   By the use of equations (16) and (17), we obtain the
torsion scalar and its derivatives in terms of $r$ as: \ba \A \A
T(r)=-\frac{4\left(rA'[(D-1)A-BC]+rCC'-\frac{1}{2}[(D-1)A-C(B-1)][(D-1)A-C(B+1)]\right)}{r^2(AD-BC)^2},
\nonu
\A \A \nonu
\A \A  \textrm{where} \qquad A'=\frac{\partial A(r)}{\partial r},
\qquad B'=\frac{\partial B(r)}{\partial r}, \qquad C'=\frac{\partial
C(r)}{\partial r} \qquad \textrm{and} \qquad D'=\frac{\partial
D(r)}{\partial r}\nonu
\A \A \nonu
\A \A T'(r)=\frac{\partial T(r)}{\partial r}=-\frac{4}{r^3(AD-BC)^3}\Biggl(r^2(AD-BC)[\{(D-1)A-BC\}A''+CC'']\nonu
\A\A-r^2\{(D-1)DA-BC(D+1)\}A'^2+4rA'\Biggl[rC'\{B(D-2)A-C(B^2+2D)\}\nonu
\A \A-r\{AD'-CB'\}[(D-2)A-BC]-D\{(D-1)A^2-ABC+C^2\}\Biggr]+r^2(AD+BC)C'^2\nonu
\A \A -2rC'\Biggl[rACD'-rC^2B'-\frac{B}{2}\{(D-1)A^2-ABC+C^2\}\Biggr]-r[(D-1)A^2-ABC+C^2](AD'+CB')\nonu
\A \A +(AD-BC)[(D-1)A-(B-1)C][(D-1)A-(B+1)C]\Biggr).\ea The equations of motion (20) can be rewritten in the form \ba \A \A
4\pi{\cal T}_0^0=-\frac{f_{TT}T'[(D-1)A^2-ABC+C^2]}{r(AD-BC)^2}
-\frac{f_{T}}{r^2(AD-BC)^3}\Biggl(rA'\{A(D-1)(DA-2BC)+C^2(B^2-D)\}\nonu
\A \A -rC'(A^2B+BC^2-2ACD)+r(AD'-CB')(A^2-C^2)+(AD-BC)[(D-1)A^2-ABC+C^2]\Biggr)
+\frac{f}{4}\,\nonu
\A \A \ea \be 4\pi{\cal
T}_1^0=\frac{4f_{TT}T'}{r(AD-BC)^2}-\frac{f_{T}[AB-DC](DA'-CB'+AD'-BC')}{r(AD-BC)^3},\ee
\be 4\pi{\cal T}_1^1=-\frac{f_{T}
\{C^2-ABC+(D-1)A^2+2rCC'+r[(D-2)A-BC]A'
\}}{r^2(AD-BC)^2}+\frac{f}{4}\,\ee \ba \A \A 4\pi{\cal
T}_2^2=4\pi{\cal
T}_3^3=\frac{f_{TT}T'}{2r(AD-BC)^2}+\frac{f_{T}}{2r^2(AD-BC)^3}\Biggl(r^2(AD-BC)[AA''-CC'']-r^2BC
A'^2\nonu
\A \A-rA'\Biggl[rC'\{BA+CD\}-rA\{AD'-CB'\}+A^2D(-2D+3) -
4BCA(1-D)-(D+2B^2)C^2\Biggr]\nonu
\A \A+r^2ADC'^2-rC'\Biggl(rACD'+r C^2B' + A^2B-4ACD-3
BC^2\Biggr)+r(AD'-CB')(A^2-C^2)\nonu
\A \A -(AD-BC)[(D-1)A-(B-1)C][(D-1)A-(B+1)C]\Biggr)+\frac{f}{4}.\ea
From equations (22)-(26), it is clear that $AD\neq BC.$

To find an exact solution of the equations (23)-(26), we impose the
following constraints:  \ba \A \A T'=0,\nonu
\A \A DA'-CB'+AD'-BC'=0,\nonu
\A \A C^2-ABC+(D-1)A^2+2rCC'+r[(D-2)A-BC]A'=0,\nonu
\A \A rA'\{A(D-1)(DA-2BC)+C^2(B^2-D)\}
-rC'(A^2B+BC^2-2ACD)+r(AD'-CB')(A^2-C^2)\nonu
\A \A +(AD-BC)[(D-1)A^2-ABC+C^2]=0.\ea As a consequence of Eq. (27), we get \ba \A \A
\Biggl(r^2(AD-BC)[AA''-CC'']-r^2BC
A'^2-rA'\Biggl[rC'\{BA+CD\}-rA\{AD'-CB'\}+A^2D(-2D+3) \nonu
\A \A-
4BCA(1-D)-(D+2B^2)C^2\Biggr]+r^2ADC'^2-rC'\Biggl(rACD'+r C^2B' + A^2B-4ACD-3
BC^2\Biggr)\nonu
\A \A
+r(AD'-CB')(A^2-C^2)-(AD-BC)[(D-1)A-(B-1)C][(D-1)A-(B+1)C]\Biggr),\ea
which does not represent any new constraint. Equations (27) are four
differential equations in four unknown functions $A$, $B$, $C$ and
$D$. The only solution of these equations is the following \be
A=1-\frac{c_1}{r}, \qquad B=\frac{c_1}{r(1-\frac{c_1}{r})}, \qquad
C=c_1/r, \qquad D=\frac{1-\frac{c_1}{r}}{1-\frac{2c_1}{r}},\ee where
$c_1$ is a constant of integration. Substituting from (29)  into
equation (22) we get a vanishing value of the scalar torsion which
satisfy the second equation of equations  (22). Therefor, solution
(29) is an exact vacuum solution of equations (23)-(26), provided
that \be f(0)=0,
 \qquad f_T(0)\neq0, \qquad f_{TT}\neq0.\ee
 To understand the physical meaning of the constant of integration  appearing in
 solution (29), $c_1$,  we are going to discuss the physics related to this solution by
 calculating the energy associated with the
tetrad  field (21) after using solution (29).

\newsection{Total Energy }

In this section, we use the solution (29) to calculate the total
energy using the translational momentum given by equation  (12). The
coframe of this solution, i.e., $  {\vartheta}^{\delta}=\left(
{h^\delta}_i \right) dx^i$,   has the form
 \ba
{\vartheta}^{\hat{0}}\A=\A Bdr+Adt, \qquad
{\vartheta}^{\hat{1}}=\sin\theta\cos\phi[D dr+Cdt]+r\cos\theta\cos\phi d\theta-r\sin\theta\sin\phi d\phi , \nonu
{\vartheta}^{\hat{2}}\A=\A \sin\theta\sin\phi[D dr+Cdt]+r\cos\theta\sin\phi d\theta+r\sin\theta\cos\phi d\phi , \qquad
{\vartheta}^{\hat{3}}= \cos\theta[D dr+Cdt]-r\sin\theta d\theta.\nonu
\A \A  \ea

The coframe (31) has the following  non-vanishing components of the
Riemannian connection \ba
 {\widetilde{\Gamma}_{\hat{1}}}^{\hat{0}}\A=\A \frac{\{[Bdr+Adt]A'-[Ddr+Cdt]C'\}\sin\theta\cos\phi+C[\sin\theta\sin\phi d\phi-\cos\theta\cos\phi d\theta]}{AD-CB}, \nonu
 {\widetilde{\Gamma}_{\hat{2}}}^{\hat{0}}\A=\A \frac{\{[Bdr+Adt]A'-[Ddr+Cdt]C'\}\sin\theta\sin\phi-C[\sin\theta\cos\phi d\phi+\cos\theta\sin\phi d\theta]}{AD-CB}, \nonu
 {\widetilde{\Gamma}_{\hat{3}}}^{\hat{0}}\A=\A \frac{\{[Bdr+Adt]A'-[Ddr+Cdt]C'\}\cos\theta+C\sin\theta d\theta]}{AD-CB}, \nonu
{\widetilde{\Gamma}_{\hat{1}}}^{\hat{2}} \A=\A\frac{[AD-BC-A]\sin^2\theta d\phi
}{AD-CB}, \qquad {\widetilde{\Gamma}_{\hat{1}}}^{\hat{3}}=\frac{[AD-BC-A][\sin\theta\sin\phi\cos\theta d\phi-\cos\phi d\theta]
}{AD-CB},\nonu
  {\widetilde{\Gamma}_{\hat{2}}}^{\hat{3}}\A=\A -\frac{[AD-BC-A][\sin\theta\cos\phi\cos\theta d\phi+\sin\phi d\theta]
}{AD-CB}.\ea
 The non-vanishing components of the superpotential 2-form  are thus\ba
\widetilde{H}_{\hat{0}} \A=\A \frac{c_1\sin \theta}{4\pi}
 (d\theta\wedge d\phi),\qquad \widetilde{H}_{\hat{1}} = -\frac{c_1\sin^2 \theta\cos\phi}{4\pi}
 (d\theta\wedge d\phi),\qquad \widetilde{H}_{\hat{2}} =-\frac{c_1\sin^2 \theta\sin\phi}{4\pi}
 (d\theta\wedge d\phi),\nonu
 \widetilde{H}_{\hat{3}} \A=\A-\frac{c_1\sin \theta\cos\theta}{4\pi}
 (d\theta\wedge d\phi).\ea Computing the total energy at a fixed time in the 3-space
with a spatial boundary 2-dimensional surface $\partial S =\{r = R,
\theta,\phi\}$,  we obtain  \be \widetilde{E}=\int_{\partial S}
\widetilde{H}_{\hat{0}}=c_1,\ee and the spatial momentum  \be
\widetilde{p_\alpha}=\int_{\partial S}
\widetilde{H}_{\hat{\alpha}}=0, \qquad \alpha=1,2,3.\ee Equation
(34) shows in a clear way that the constant of integration $c_1$ is
related to the gravitational mass $M$ \cite{MTW}.
\newpage
\newsection{Main results and discussion}

  The $f(T)$ gravitational theory has been considered in the vacuum case. The  equations of motion   have been
applied to a non-diagonal  spherically  symmetric tetrad field
having four unknown functions of the redial coordinate. Four
nonlinear differential equations are obtained. To  solve these
differential equations, we have imposed four constraints (in four
unknown functions). The only solution that is compatible with these
constraints contains  one constant of integration. {\it Therefore,
an  exact vacuum spherically symmetric solution to the field
equations of $f(T)$ gravitational theory has been derived}. This
solution has a vanishing scalar torsion and is satisfying the
equations of motion of $f(T)$,  provided that
 equation  (30) holds.  To understand  the
physical meaning  of the constant of integration, we calculate its
associated energy. It has been shown that it is related to the
gravitational mass.

To understand the construction of the derived solution, let us
rewrite it in the following form:
 \ba \A \A \left( {h^\mu}_i \right)= \left({\Lambda^\mu}_\nu\right)_1\left({\Lambda^\nu}_\lambda\right)_2
  \left( {h^\lambda}_i \right)_d,\qquad \textrm {where} \nonu
\A \A \left({\Lambda^\mu}_\nu\right)_1=\left( \matrix{ 1 &  0 & 0 & 0
\vspace{3mm} \cr  0  &  \sin\theta \cos\phi &  \cos\theta \cos\phi &
- \sin\phi \vspace{3mm} \cr 0  & \sin \theta \sin \phi& \cos\theta
\sin\phi & \cos\phi \vspace{3mm} \cr 0  & \cos\theta & -\sin\theta &
0 \cr }\right)\;, \qquad \left({\Lambda^\mu}_\nu\right)_2=\left( \matrix{
\displaystyle\frac{1-\frac{M}{r}}{1-\frac{2M}{r}} &
\displaystyle\frac{M}{r(1-\frac{2M}{r})} & 0 & 0 \vspace{3mm} \cr
\displaystyle\frac{M}{r(1-\frac{2M}{r})} &
\displaystyle\frac{1-\frac{M}{r}}{1-\frac{2M}{r}} &0& 0 \vspace{3mm}
\cr 0  & 1&0 &0 \vspace{3mm} \cr 0  & 0 & 0 & 1 \cr }\right)\;,\nonu
\A \A \left( {h^\mu}_i \right)_d=\left( \matrix{
\sqrt{1-\frac{2M}{r}} & 0 & 0 & 0 \vspace{3mm} \cr 0 &
\displaystyle\frac{1}{\sqrt{1-\frac{2M}{r}}} &0& 0 \vspace{3mm} \cr
0  & 0&r &0 \vspace{3mm} \cr 0  & 0 & 0 & r\sin\theta \cr
}\right)\;.\ea The matrices  (36) show that the exact solution
consists of a diagonal
 solution given by $\left( {h^\mu}_i \right)_d$ and two local
 Lorentz transformations, i.e., $\left({\Lambda^\mu}_\nu\right)_1$ and
$\left({\Lambda^\mu}_\nu\right)_2$. The
 tetrad $\left({\Lambda^\mu}_\nu\right)_1\left( {h^\nu}_i \right)_d$
 has been applied to $f(T)$ theory \cite {Ncpl,DRH} and no exact solution has been obtained
 but  only an asymptotic one, that represents the Schwarzschild-Ads and traversable wormhole
 solution. The extra local Lorentz transformation $\left({\Lambda^\mu}_\nu\right)_2$
 plays a key role in adjusting the solution to be exact  to
 the $f(T)$ gravitational theory.

 Maluf et al. \cite{MFU} have used a tetrad that can be decomposed into two
 local Lorentz transformations and a diagonal tetrad. This tetrad
 has given the Schwarzschild spacetime and it has been shown that the energy
 content related to this tetrad is vanishing. Maluf et al. \cite{MFU} have discussed this result as
  a freely falling test body  in this
 spacetime. However, Obukhov et al.  \cite{LOP} have shown that there is inertia
 that contributes to the gravitational mass. This result produces  the vanishing energy.
 Here, in this study, we have obtained a solution similar to  that
 studied by Maluf et al. \cite{MFU} and Lucas et al. \cite{LOP}.  Our$'$s solution is   similar in the sense that it consists
 of two local Lorentz transformations and a diagonal tetrad.
 There is no inertia contributing to the physics as shown by equation  (34).

\end{document}